\documentclass[11pt]{article}
\usepackage{hyperref}
\usepackage{latexsym}
\def \dd{{\rm d}}

\def \II{{\rm I}}

\begin{document}
\newcommand{\ggg}[2]{\mathbf{g}{\shortstack{\scriptsize $#1#2$\tiny\vspace{-2pt}\\
$\sf v$}}}
\newcommand{\ggd}[2]{g_{\shortstack{\rule{0pt}{0pt}\\
\scriptsize $#1#2$\tiny\vspace{-2pt}\\ $\sf v$}}}
\newcommand{\Rgd}[2]{R_{\shortstack{\rule{0pt}{0pt}\\
\scriptsize $#1#2$\tiny\vspace{-2pt}\\ $\sf v$}}}
\newcommand{\xxs}[2]{{#1_{\hspace{-5pt}\shortstack{\rule{0pt}{4pt}\\
\scriptsize #2}}}{}}
\newcommand{\xgd}[3]{#1_{\shortstack{\rule{0pt}{0pt}\\
\scriptsize $#2#3$\tiny\vspace{-2pt}\\ $\sf v$}}}
\newcommand{\ddv}[2]{_{\shortstack{\scriptsize $#1#2$\tiny\vspace{-3pt}\\ $\sf v$}}}
\newcommand{\uuv}[2]{{\shortstack{\scriptsize $#1#2$\tiny\vspace{-2pt}\\ $\sf v$}}}
\title{Hans-J\"urgen Treder and the discovery of
confinement in Einstein's unified field theory}
\author{{Salvatore Antoci}\\\bigskip \footnotesize{Dipartimento di Fisica
``A. Volta'' and IPCF of CNR, Pavia, Italia}\\
{Dierck Ekkehard Liebscher}\\
\footnotesize{Astrophysikalisches Institut Potsdam, Potsdam,
Deutschland}}
\date{}
\maketitle
\begin{abstract}
In the year 1957, when interest in Einstein's unified field theory
was fading away for lack of understanding of its physical content,
Treder performed a momentous critical analysis of the possible
definitions of the electric four-current in the theory. As an
outcome of this scrutiny he was able to prove by the E.I.H. method
that properly defined point charges, appended at the right-hand
side of the field equation $R_{\shortstack{\scriptsize
$[\mu\nu$\tiny\vspace{-3pt}\\ $~\sf v$} \shortstack{\scriptsize
$,\lambda]$ \\ \rule{0pt}{2pt}}}=0$, interact mutually with
Coulomb-like forces, provided that a mutual force independent of
distance is present too. This unwanted, but unavoidable addition,
could not but lay further disbelief on the efforts initiated by
Einstein and Schr\"odinger one decade earlier. However in 1980
Treder himself recalled that the potential $\varphi=a/r+cr$, found
by him in 1957, was the one used by particle physicists to account
phenomenologically for the spectrum of bound quark systems like
mesons. Exact solutions have later confirmed beyond any doubt that
Einstein's unified field theory does account in a simple way,
already in classical form, for the confinement of pole charges
defined by the four-current first availed of by Treder.\par In the
present paper it is proposed, ad memoriam, a thorough recollection
of the article published by Treder in 1957, showing the way kept
by him to find what would have been later recognized as
confinement in Einstein's unified field theory.
\end{abstract}
\section{Introduction}
In the year 1957, one decade had elapsed since Einstein had
resumed his attempt\cite{Einstein1925}, first formulated already
in 1925, to encompass both gravitation and electromagnetism in a
generalization of his theory of 1915 based on a nonsymmetric
fundamental tensor $g_{\mu\nu}$, and on a nonsymmetric affine
connection $\Gamma^{\lambda}_{\mu\nu}$. One decade too had gone by
after Schr\"odinger, by starting from a purely affine approach,
had come to announce\cite{Schroedinger1947} to have eventually
reached ``the final affine field laws'', that would encompass too
both electromagnetism and gravitation in a geometric formulation,
very similar to the one proposed by Einstein.\par However, despite
the intense work done by many relativists and geometers to
understand both the mathematical structure and the physical
content of what was appropriately called the
Einstein-Schr\"odinger theory, the perspective for this sort of
endeavour was not, in 1957, as promising as it had appeared one
decade earlier. The formal simplicity of the sets of equations
proposed both by Einstein and by Schr\"odinger had not yet found a
counterpart in an equally simple and satisfactory physical
interpretation. Already in the 1954/55 report to the Dublin
Institute for Advanced Studies, a disappointed Erwin Schr\"odinger
had written: ``It is a disconcerting situation that ten years
endeavour of competent theorists has not yielded even a plausible
glimpse of Coulomb's law."\cite{Hittmair1987}.\par In the very
year 1957 H. Treder, a collaborator of A. Papapetrou in Berlin,
published in Annalen der Physik a paper, where a critical scrutiny
of the definitions of the electric charge-current density
admissible in Einstein's unified field theory is
performed\cite{Treder1957} for the first time. As a consequence of
his analysis of this crucial issue, Treder could show that the
negative outcome for the electric force, found both by
Infeld\cite{Infeld1950} and by Callaway\cite{Callaway1953}, by
availing of the weak field approximation for solving the equations
of motion by the E.I.H. method\cite{EIH1938,EI1949}, depended on
the choice of the definition of the electric four-current done by
those authors. If a different choice is made, allowed for too
according to Treder's analysis, the equations of motion stemming
from the weak field approximation prove that Einstein's unified
field theory does admit of non gravitational forces between
charged point particles. But a surprising result is found with
Treder's choice of the electric four-current too, for Einstein's
theory does not provide in this way a pure Coulomb force between
two point charges. A force independent of distance is also
present, that cannot be made to vanish by any choice of the
constants, because the presence of the latter force is mandatory
for the very existence of the Coulomb term.\par

Already in his paper of 1957, Treder expressed some doubt about
the electromagnetic meaning of his finding. Indeed, when the two
charged particles are sufficiently far away from each other, the
component of the force that is independent of distance will
inescapably become the prevailing one, no matter how weak it may
be chosen to be.\par Scope of the present paper is to reconsider
Treder's finding of 1957, based on approximate calculations, in
full detail. It will be reminded too that in 1980 Treder
himself\cite{Treder1980} interpreted his earlier finding as
proving that, in the Hermitian version\cite{Einstein1948} of
Einstein's theory, pole point charges of unlike sign, provided by
the four-current first considered by him\cite{Treder1957}, attract
mutually with a force independent of distance, hence they are
permanently confined entities, like the quarks of chromodynamics
are presumed to be.\par It will be reminded eventually that exact
solutions to the field equations of Einstein's unified field
theory belonging to a class found\cite{Antoci1987} in 1987 confirm
with exact arguments \cite{Antoci1984, ALM2006} the existence of
confinement in Einstein's unified field theory.\par

\section{Treder's definition of the charge-current in Einstein's unified field theory}

It is quite interesting to examine the logical thread followed by
Treder in choosing his definition of the ``electric'' 4-current
density. It is evident that in 1957, given the problematic
condition of the theory, he feels the need to confront the issue
afresh, without being encumbered by prejudices, in particular by
the authoritative a priori stipulation, upheld both by
Einstein\cite{Einstein1949} and by
Schr\"odinger\cite{Schroedinger1947}, according to which, since
the new theory did represent the field-theoretical completion of
the theory of 1915, no phenomenological sources had to be appended
at the right-hand sides of its field equations. For Treder, it is
the so-called $+-$ relation that plays a crucial guiding r\^ole.
According to him, it is evident that, since the equation
$g_{\shortstack{\scriptsize $\mu\nu$\vspace{-2pt}\\ {\tiny $+-$}}
   \shortstack{\scriptsize $;\lambda$ \\ \rule{0pt}{2pt}}}=0$
provides the definition of the affine connection
$\Gamma^{\lambda}_{\mu\nu}$ in terms of the fundamental tensor
$g_{\mu\nu}$, it needs to be satisfied everywhere.\par As a
consequence of this stipulation, also the electromagnetic looking
equation $\ggg{\mu}{\nu},_{\nu}=0$, that
stems\cite{Schroedinger1947} from the previous defining equation
for $\Gamma^{\lambda}_{\mu\nu}$, needs to be satisfied everywhere.
Therefore it is impossible to interpret the latter equation as
representing, in Einstein's unified field theory, the first group
of Maxwell's equations. In Einstein's theory, $\ggg{\sigma}{\tau}$
``must be the antisymmetric tensor density dual to the
electromagnetic field strength''. But this momentous recognition
is of scarce help in deciding what 4-vector represents the
electric four-current, `` because in the unified field theory
field strength and induction are not necessarily connected through
the relation which we know from Maxwell's theory''.\par In order
to confront the issue, Treder carefully examines, in part {\bf II}
of his paper, what suggestions come from the solutions of the weak
field, first order approximation of the field equation
$R_{\shortstack{\scriptsize $[\mu\nu$\tiny\vspace{-3pt}\\ $~\sf
v$} \shortstack{\scriptsize $,\lambda]$ \\ \rule{0pt}{2pt}}}=0$,
because, ``when the unified field theory is expected to have also
a macroscopic meaning, it must be required that it allows to
describe the existence of pointlike charges in the classical
vacuum in the lowest approximation at least, for weak fields''.
According to Treder, in order to solve the issue of the definition
of the electric four-current, there is therefore merit in studying
the spherically symmetric, static solution of the weak field
approximation of $R_{\shortstack{\scriptsize
$[\mu\nu$\tiny\vspace{-3pt}\\ $~\sf v$} \shortstack{\scriptsize
$,\lambda]$ \\ \rule{0pt}{2pt}}}=0$. Since,
$\ggg{\mu}{\nu},_{\nu}=0$ must be fulfilled everywhere, the first
order approximation of $\ggd{\mu}{\nu}$ must be the dual of the
curl of a four-vector:
$$
\xxs{g}{1}\ddv{\mu}{\nu}=\frac{1}{2}\varepsilon_{\mu\nu\sigma\tau}
\left(\varphi_{\underline\tau,\underline\sigma}
-\varphi_{\underline\sigma,\underline\tau}\right),
$$
and in the Lorentz gauge $\xxs{R}{1}_{\shortstack{\scriptsize
$[\mu\nu$\tiny\vspace{-3pt}\\ $~\sf v$} \shortstack{\scriptsize
$,\lambda]$ \\ \rule{0pt}{2pt}}}=0$ specialises to
$$
\Box^2\,\varphi^{\sigma}=0.
$$
In the static case $\varphi^{\lambda}=(0,0,0,\varphi)$, and the
latter equation specialises further to
$$
\Delta\Delta\varphi=0,
$$
whose general, spherically symmetric solution reads\footnote{Biharmonic
equations are to be expected in theories with quadratic Lagrangians.
It is
remarkable that the same weak field expression of a potential
is found too, under suitable conditions,
in the framework of Poincar\'e gauge field
theory.
For instance, with the purpose and interpretation
of confinement this was considered in
the paper ``Short-range confining
component in a quadratic Poincar\'e  gauge theory of
gravitation''\cite{HNNV1978}.}
$$
\varphi=\frac{a}{r}+b+cr+dr^2.
$$
After dropping the term $dr^2$, that leads to the divergent
behaviour of ${g_{\hspace{-5pt}\shortstack{\rule{0pt}{4pt}\\
\scriptsize 1}}}{}_{\mu\nu}$ for $r=\infty$, and the unessential
constant $b$, $\varphi$ takes the paradigmatic form
$$
\varphi=\frac{a}{r}+cr.
$$
Treder then looks for the charge density definitions that are
compatible with this form of $\varphi$ in the previously specified
sense, namely, he looks for the $\delta$ functions that can be
generated through either single or double application of the
Laplace operator $\Delta$ to $\varphi$, and seeks to what exact
charge-current density definitions they shall correspond as
particular approximate, weak field cases. He notices that by
applying the Laplace operator once to the first term of $\varphi$
one gets
$$
\Delta\left(\frac{a}{r}\right)=-4\pi a\delta(\mathbf r)
$$
i.e. a $\delta$ source that is a particular static, first order
approximation of the source term $s_{\mu\nu\lambda}$ occurring in
the general four-current definition
$$
g_{\shortstack{\scriptsize $[\mu\nu$\tiny\vspace{-3pt}\\ $~\sf v$}
        \shortstack{\scriptsize $,\lambda]$ \\ \rule{0pt}{2pt}}}
         \equiv -s_{\mu\nu\lambda}.
$$
Against this option, however, Treder raises the objection that it
does not allow for a free choice of the charge-current density,
like it happens instead in Maxwell's theory, because, due to the
field equation $R_{\shortstack{\scriptsize
$[\mu\nu$\tiny\vspace{-3pt}\\ $~\sf v$} \shortstack{\scriptsize
$,\lambda]$ \\ \rule{0pt}{2pt}}}=0$, the charge-current density
defined in this way is constrained to fulfill, in the weak field
approximation, a differential, d'Alembert equation:
$$
\Box\xxs{s}{1}_{\mu\nu\lambda}=0.
$$
By applying twice the Laplace operator to the second term of
$\varphi$ one gets again a $\delta$ function:
$$
\Delta\Delta(cr)=-8\pi c\delta(\mathbf r).
$$
Treder notices that this particular $\delta$ charge density is a
weak field, static instance of a charge-current density
$s_{\mu\nu\lambda}$ defined by
$$
R_{\shortstack{\scriptsize $[\mu\nu$\tiny\vspace{-3pt}\\ $~\sf v$}
        \shortstack{\scriptsize $,\lambda]$ \\ \rule{0pt}{2pt}}}
\equiv\frac{1}{2}s_{\mu\nu\lambda},
$$
i.e. of a charge-current density appended in a phenomenological
way at the right-hand side of the field equation
$R_{\shortstack{\scriptsize $[\mu\nu$\tiny\vspace{-3pt}\\ $~\sf
v$} \shortstack{\scriptsize $,\lambda]$ \\ \rule{0pt}{2pt}}}=0$.
As such, this four-current density can be assigned at will
(subject to the conservation law) like it occurs in Maxwell's
theory. When both terms of $\varphi$ are considered, the double
application of the Laplacian leads instead to the point source
expression
$$
\Delta\Delta\varphi= -4\pi \left(a\Delta\delta(\mathbf
r)+2c\delta(\mathbf r)\right).
$$
This is the structure of each of the $n$ point charges that Treder
introduces in his derivation of the equations of motion in the
weak field, first order approximation of the E.I.H. method
performed in {\bf III}, after having assumed that, for any charge
$\II$, the ratio $c_\II/a_\II$ is a universal constant $\tau$.

\section{Treder's discovery of confinement in Einstein's unified field theory}

After the momentous assumptions done in {\bf II}, deriving in {\bf
III}, by the E.I.H. method, the equations of motion for $n$
charged particles in the weak field, slow motion approximation is
for Treder a straightforward, routine move, done in the footsteps
of Infeld\cite{Infeld1950}. It leads however to a highly
perplexing end result. Like Treder, let us consider for simplicity
the case $n=2$, when the static potential $\varphi$ comes to read
$$
\varphi=\varphi_{\II}+\varphi_{\II\II}
=\frac{a_{\II}}{r_{\II}}+c_{\II}r_{\II}
+\frac{a_{\II\II}}{r_{\II\II}}+c_{\II\II}r_{\II\II}
$$
and, to the required order of approximation, although with some
inappropriateness in the language, one may assert that the
Cartesian components of the ``electric'' force that the field of a
pointlike charge $\II\II$ exerts on a pointlike charged particle
$\II$ is given by
\begin{eqnarray}\nonumber
\xxs{\stackrel{\II}{\mathcal{L}}}{4}_i
\equiv\frac{1}{4\pi}\oint\limits_\II^0
2~\xxs{L}{4}_{ik}\stackrel{\II}{n}^k\dd\stackrel{\II}{S}\\\nonumber
=2c_\II a_{\II\II}\frac{\xi_i}{\varrho^3} +2a_\II
c_{\II\II}\frac{\xi_i}{\varrho^3} -2c_\II
c_{\II\II}\frac{\xi_i}{\varrho}\\\nonumber
\xi^i=x^i_\II-x^i_{\II\II} \ {\rm and} \ \varrho^2=\xi^s\xi^s,
\end{eqnarray}
when the integral is extended to a closed surface surrounding only
particle $\II$. When this expression, found by Treder as a direct
outcome of his pondered choice of the definition of the
four-current in Einstein's unified field theory, appeared in
print\cite{Treder1957}, it was not new. It had been written
already by V. V. Narlikar and B. R. Rao in their paper of 1956,
entitled ``The equations of motion of particles in the unified
field theory of Einstein (1953)''\cite{Narlikar Rao1956}. However
we feel entitled to attribute only to Treder the correct
interpretation of this surprising result, and to continue its
analysis along the line drawn in his article of 1957, because the
interpretation considered by Narlikar and Rao is instead based on
an untenable assumption. For these authors, the four-current
responsible for the above written force is proportional to
$g_{\shortstack{\scriptsize $[\mu\nu$\tiny\vspace{-3pt}\\ $~\sf
v$}
        \shortstack{\scriptsize $,\lambda]$ \\ \rule{0pt}{2pt}}}$, hence the
corresponding charges are by no means pointlike, but diffused in
the whole space and overlapping. It is obvious that point
particles are instead needed to make sense of an E.I.H.
calculation.\par However, although Treder's choice of the
four-current leads in the present case to pointlike charges, i.e.
his E.I.H. calculation is conceptually faultless, the ``electric''
interpretation of the resulting force soon arose perplexity.
F.A.E. Pirani, when commenting\cite{Pirani1957} Treder's paper for
the ``Mathematical Reviews'', wrote:
\begin{quotation}
 The author proposes a new definition of charge-current in Einstein's
 ``weak'' non-symmetric unified theory [The meaning of relativity, 3rd ed.,
 revised, Princeton, 1950, Appendix II]. In the lowest approximation
 he obtains the Coulomb force between point charges, but also,
 unfortunately, an additional force independent of distance.
\end{quotation}
As noted by Treder, only if one makes the additional assumption
that the ratio $c_\II/a_\II,~i=1,\cdot\cdot\cdot,~n$ is a
universal constant $\tau$ does one get a law of force that
approximates the ordinary Coulomb law, as long as the inequality
$$
\tau\ll \frac{1}{\varrho^2}
$$
is satisfied. But of course, the force independent of distance
cannot be hidden out: it will always become the prevailing one
when the charged particles are posited farther and farther away
from each other. Therefore, the ``electric'' interpretation of the
result, although it found its adherents, e.g. in\cite{Clauser1958,
Johnson1972}, was never considered to be a satisfactory one, not
even by Treder himself at the very moment of its finding, as it
transpires from the concluding remarks in {\bf V}.\par It is
evident that in 1957 a force independent of distance between point
charges could not be thought to be of much use in theoretical
physics. Therefore the very existence of such a force in
Einstein's unified field theory, so keenly brought into evidence
by Treder, could not but help laying further discredit on the
theoretical endeavour inaugurated by Einstein and Schr\"odinger
one decade earlier.\par However, what ideas are of interest to
theoreticians change with the lapse of time and, as mentioned in
the Introduction, in 1980 Treder\cite{Treder1980} might well
wonder whether his early finding could not be reinterpreted as the
evidence that Einstein's theory allows, already in classical form,
for the confinement of quarks, i.e. it can account for both the
strong and the gravitational force in a unified way.
Phenomenological potential models introduced at the
time\cite{Eichten1978, Eichten1980} used in fact a linear
combination of a Coulomb and of a linear radial potential, just
like the one found by Treder in 1957, to account satisfactorily
for the spectroscopy of hadrons. But, one should ask: if
Einstein's theory allows for a unified description of both the
strong and the gravitational interaction, where must one look for
electromagnetism in the theory? What entity represents, in the
theory, the long sought for electric four-current?\par

\section{What the exact solutions have to say}

After the finding, in 1987, of a class of exact solutions of the
Einstein-Schr\"odinger equations intrinsically depending on three
coordinates\cite{Antoci1987} it was noticed, from the study of
particular solutions, that perhaps Treder's injunction, that both
the
equations $g_{\shortstack{\scriptsize $\mu\nu$\vspace{-2pt}\\
{\tiny $+-$}}
   \shortstack{\scriptsize $;\lambda$ \\ \rule{0pt}{2pt}}}=0$
   and $\ggg{\mu}{\nu},_{\nu}=0$ have to be
satisfied everywhere, is too restrictive, thereby leading to a
loss of valuable physical content of the theory. In 1978
Borchsenius\cite{Borchsenius1978} had shown how to obtain that a
phenomenological four-current may appear at the right-hand sides
of the two equations just mentioned above without destroying the
invariance of the theory under transposition. Therefore, in the
footsteps of the successful phenomenological completion of the
general relativity of 1915, the way was open for interpreting
Einstein's theory of the nonsymmetric field as a theory admitting
both a symmetric energy tensor $T_{\mu\nu}$ and two distinct,
conserved four-currents $j^{\varrho}$ and $K_{\mu\nu\lambda}$ like
phenomenological sources\cite{Antoci1991}. Its field equations,
that reduce to the original ones wherever sources are absent, then
read:
$$
\mathbf{g}^{\mu\nu}_{~,\lambda}+\mathbf{g}^{\sigma\nu}\Gamma^\mu_{\sigma\lambda}+
\mathbf{g}^{\mu\sigma}\Gamma^\nu_{\lambda\sigma}
-\mathbf{g}^{\mu\nu}\Gamma^\tau_{\underline{\lambda\tau}}
=\frac{4\pi}{3}(\mathbf{j}^\mu\delta^\nu_\lambda-\mathbf{j}^\nu\delta^\mu_\lambda),
$$
$$
\ggg{\varrho}{\sigma},_{\sigma}={4\pi}\mathbf{j}^{\varrho},
$$
$$
\bar{R}_{\underline{\mu\nu}}(\Gamma)=8\pi(T_{\mu\nu}
-\frac{1}{2}s_{\mu\nu}s^{\varrho\sigma}T_{\varrho\sigma}),
$$
$$
\bar R_{\shortstack{\scriptsize $[\mu\nu$\tiny\vspace{-3pt}\\
$~\sf v$} \shortstack{\scriptsize $,\lambda]$ \\
\rule{0pt}{2pt}}}=8\pi K_{\mu\nu\lambda},
$$
where, like in Treder's paper\cite{Treder1957}, $s_{\mu\nu}$ is
the metric tensor defined by
Kur\c{s}u\-no\u{g}lu\cite{Kursunoglu1952} and
H\'ely\cite{Hely1954} as
$$
s^{\mu\nu}=\sqrt{\frac{g}{s}}g^{\underline{\mu\nu}}~, \ \
s_{\sigma\tau}s^{\mu\tau}=\delta^{\mu}_{\sigma}.
$$
$\bar{R}_{\mu\nu}$ is the symmetrised Ricci tensor of Borchsenius:
$$
\bar{R}_{\mu\nu}(\Gamma)=\Gamma^\varrho_{\mu\nu,\varrho}
-\frac{1}{2}\left(\Gamma^\varrho_{\mu\varrho,\nu}
+\Gamma^\varrho_{\nu\varrho,\mu}\right)
-\Gamma^\alpha_{\mu\varrho}\Gamma^\varrho_{\alpha\nu}
+\Gamma^\alpha_{\mu\nu}\Gamma^\varrho_{\alpha\varrho},
$$
that reduces to the plain one wherever $\mathbf{j}^{\varrho}$
vanishes. With these definitions, and with the semicolon ``;''
standing for the covariant differentiation performed with the
Christoffel symbols built with $s_{\mu\nu}$, the contracted
Bianchi identities of the theory come to read
$$
\mathbf{T}^{\lambda\sigma}_{~;\sigma}=\frac{1}{2}s^{\lambda\varrho}
\left(\mathbf{j}^\tau\bar\Rgd{\varrho}{\tau}(\Gamma)
+K_{\tau\varrho\sigma}\ggg{\sigma}{\tau}\right), \ \
\mathbf{T}^{\mu\nu}=\sqrt{-s}s^{\mu\varrho}s^{\nu\sigma}T_{\varrho\sigma},
$$
a perspicuous enough writing.\par To the previously mentioned
class of solutions belongs a particular exact solution that is
static and endowed with pole charges built with the current
$K_{\tau\varrho\sigma}$. Its details are given
elsewhere\cite{Antoci1984,ALM2006} and will not be repeated here.
Suffice it to say that the solution confirms beyond any possible
doubt what the approximate result found by Treder in 1957 already
said, i.e. that Einstein's unified field theory, when complemented
with the phenomenological four-current $K_{\tau\varrho\sigma}$,
allows describing point charges interacting mutually with forces
independent of distance. In the Hermitian version of the theory
two charges of unlike sign mutually attract, hence are permanently
confined entities. As far as exact solutions are concerned, the
theory therefore provides examples both of gravitating
bodies\cite{BW22} and of bodies interacting like quarks are
expected to do.\par But to the same class belongs another exact
solution\cite{ALM2005}, that is static too, and whose field
$\ggg{\mu}{\nu}$ is associated with charge density built with the
other four-current, $j^\varrho$. Since, outside the charges, the
field fulfils the field equation $\ggg{\mu}{\nu},_{\nu}=0$, while
the unsolicited equation
$$
g_{\shortstack{\scriptsize $[\mu\nu$\tiny\vspace{-3pt}\\ $~\sf v$}
        \shortstack{\scriptsize $,\lambda]$ \\ \rule{0pt}{2pt}}}=0
$$
is satisfied everywhere, one cannot help recognizing in this
solution the general electrostatic solution of Einstein's unified
field theory. Moreover if, in the adopted representative space,
one puts the charge distribution on $n$ localized, closed
two-surfaces, it is possible\cite{ALM2005} to generate, in the
metric sense, the charge distribution of $n$ pointlike,
spherically symmetric charges. This occurrence only happens when
the charges occupy mutual positions that correspond, with all the
accuracy needed to meet with the most stringent empirical results,
to the mutual positions dictated by Coulomb's law for the
equilibrium condition of $n$ pointlike charges.\par As far as the
evidence associated with a particular exact solution can go, this
result constitutes a partial, but hopefully enlightening answer to
the two questions raised at the end of the previous section, about
the presence of electromagnetism in Einstein's theory, and about
the identification of the electric four-current.\par{}

\end{document}